\documentclass{an}
\usepackage{graphicx}
\usepackage{times}
\usepackage{fancyhdr}
\begin{document}

\title{Observational characteristics of dense cores with deeply embedded young protostars}

\author{D. Stamatellos, A.P. Whitworth \and S.P. Goodwin}
\institute{
School of Physics \& Astronomy, Cardiff University, 5 The Parade, CF24 3YB, 
Cardiff, UK
}

\date{Received $<$date$>$; 
accepted $<$date$>$;
published online $<$date$>$}

\def\micron{\mu {\rm m}}
\def\degr{\hbox{$^\circ$}}
\def\arcmin{\hbox{$^\prime$}}
\def\arcsec{\hbox{$^{\prime\prime}$}}

\def\aj{AJ}%
\def\actaa{Acta Astron.}%
\def\araa{ARA\&A}%
\def\apj{ApJ}%
\def\apjl{ApJ}%
\def\apjs{ApJS}%
\def\ao{Appl.~Opt.}%
\def\apss{Ap\&SS}%
\def\aap{A\&A}%
\def\aapr{A\&A~Rev.}%
\def\aaps{A\&AS}%
\def\azh{AZh}%
\def\baas{BAAS}%
\def\bac{Bull. astr. Inst. Czechosl.}%
\def\caa{Chinese Astron. Astrophys.}%
\def\cjaa{Chinese J. Astron. Astrophys.}%
\def\icarus{Icarus}%
\def\jcap{J. Cosmology Astropart. Phys.}%
\def\jrasc{JRASC}%
\def\mnras{MNRAS}%
\def\memras{MmRAS}%
\def\na{New A}%
\def\nar{New A Rev.}%
\def\pasa{PASA}%
\def\pra{Phys.~Rev.~A}%
\def\prb{Phys.~Rev.~B}%
\def\prc{Phys.~Rev.~C}%
\def\prd{Phys.~Rev.~D}%
\def\pre{Phys.~Rev.~E}%
\def\prl{Phys.~Rev.~Lett.}%
\def\pasp{PASP}%
\def\pasj{PASJ}%
\def\qjras{QJRAS}%
\def\rmxaa{Rev. Mexicana Astron. Astrofis.}%
\def\skytel{S\&T}%
\def\solphys{Sol.~Phys.}%
\def\sovast{Soviet~Ast.}%
\def\ssr{Space~Sci.~Rev.}%
\def\zap{ZAp}%
\def\nat{Nature}%
\def\iaucirc{IAU~Circ.}%
\def\aplett{Astrophys.~Lett.}%
\def\apspr{Astrophys.~Space~Phys.~Res.}%
\def\bain{Bull.~Astron.~Inst.~Netherlands}%
\def\fcp{Fund.~Cosmic~Phys.}%
\def\gca{Geochim.~Cosmochim.~Acta}%
\def\grl{Geophys.~Res.~Lett.}%
\def\jcp{J.~Chem.~Phys.}%
\def\jgr{J.~Geophys.~Res.}%
\def\jqsrt{J.~Quant.~Spec.~Radiat.~Transf.}%
\def\memsai{Mem.~Soc.~Astron.~Italiana}%
\def\nphysa{Nucl.~Phys.~A}%
\def\physrep{Phys.~Rep.}%
\def\physscr{Phys.~Scr}%
\def\planss{Planet.~Space~Sci.}%
\def\procspie{Proc.~SPIE}%
\let\astap=\aap
\let\apjlett=\apjl
\let\apjsupp=\apjs
\let\applopt=\ao

\abstract{
Class 0 objects, which are thought to be
the youngest protostars, are identified in terms of NIR or radio 
emission and/or the presence of molecular outflows. We present combined hydrodynamic and 
radiative transfer simulations of the collapse of a star-forming molecular core,
which suggest two criteria for identifying dense cores with deeply embedded very young protostars 
that may not be observable in the NIR or radio with current telescopes.
We find that cores with protostars are relatively warm (T$>15$~K) with their
SEDs peaking at wavelengths $<170\micron$, and they tend to appear circular.
}

\correspondence{D.Stamatellos@astro.cf.ac.uk}

\maketitle

\section{Introduction}

Class 0 objects, the youngest protostars identified so far, are distinguished
from prestellar cores, their precursors, based on 2 criteria (see Andr\'e et al.~2000), 
(i) the presence of a NIR source in the centre of the core, or
(ii) the presence of compact radio emission or molecular outflows. These
criteria depend on the sensitivity of the telescope used.
Very young protostars  are difficult to observe in the NIR, because they 
are deeply embedded. Additionally,  the amount of their radio emission 
 is uncertain (see Gibb 1999), and probably very low 
(Neufeld \& Hollenbach 1996). Thus, a null detection with a specific telescope 
does not necessarily mean that a source does not exist. 
This was demonstrated recently with the core L1014 which was thought to be prestellar 
because no source was detected by IRAS, but later was suggested  to be protostellar
after a NIR source was detected by the more sensitive SPITZER 
(Young et al. 2004).  Hence it is important to consider ways in which 
the presence of young embedded protostars might be inferred indirectly. 
In this paper we present SPH simulations of the collapse of a turbulent molecular 
core, combined with 3-dimensional Monte Carlo radiative transfer simulations at 
different stages of this collapse. The simulations  predict dust temperature fields, 
SEDs, and isophotal maps of very young protostars that are deeply embedded in
their parental cores, and provide criteria 
for distinguishing between genuine prestellar cores and cores that contain 
very young protostars (Stamatellos et al. 2005). 

\newpage
\begin{table*}
\begin{center}
\small\
\caption{Model parameters}
\begin{tabular}{@{}cccccccccl} \hline
id & time (yr) & $M_\star ({\rm M}_{\sun})$ 
& $R_{\rm S}$ (AU) & $Z_{\rm S}$ (AU)& $M_{\rm S} ({\rm M}_{\sun})$ 
& $\dot{M}_\star ({\rm M}_{\sun}$/yr) 
& $L_{\rm TOT} ({\rm L}_{\sun})$ & $T_\star$ (K) 
& description \\
\hline
\texttt{t1} & $5.3\times10^{4}$ & - & - &  - & - & - & - & - & prestellar core \\
\texttt{t3} & $6.0\times10^{4}$ & 0.20  & 4.0 &  0.4 & 0.09 & $1\times10^{-5}$& 27.2 & 7650 & 
Class 0 protostar\\
\texttt{t5} & $6.9\times10^{4}$ & 0.53 & 4.0 &  0.4 & 0.01 & $4\times10^{-7}$& 2.5  & 4200 & protostar is off centre \\
\hline
\end{tabular}\label{tab:model.params}
\end{center}
\hspace{0.8cm}
\begin{tabular}{@{}lll}
$Z_{\rm S}$, $R_{\rm S}$: Smartie dimensions  &  $M_{\rm \star}$: Stellar mass&
$T_\star$: Star temperature\\
$M_{\rm S}$: Smartie mass  & $\dot{M}_\star$: Accretion rate onto the central star
&$L_{\rm TOT}$: Total star luminosity (intrinsic +accretion) \\
\end{tabular}
\end{table*}


\section{The model}

We use the SPH code {\sc Dragon} (Goodwin et al. 2004) to simulate
the dynamics of a collapsing molecular core and the Monte Carlo radiative
transfer code {\sc Phaethon} (Stamatellos \& Whitworth 2003, 2005; 
Stamatellos et al. 2004) to treat the radiative transfer within the core, at
different time-frames during its collapse.

\subsection{The collapse of a molecular cloud}

The initial conditions in the core, before the collapse, are set according to
observations (e.g. Kirk et al. 2005).
Prestellar cores have typical extents from 2000 to $>15000$~AU, and
masses from 0.05 to 10 M${_\odot}$. Their density profiles are flat in their centre
and fall off as $r^{-n}$, where $n=2-4$. These features are well
 represented by a Plummer-like density profile (Plummer 1915), 
\begin{equation} 
\rho=\frac{\rho_{_0}}{\left(1+(r/r_{_0})^2\right)^2}\,,\;\;\;\;r < r_{_{\rm B}}\,.
\end{equation}
The density profile is almost flat for $r < r_{_0}$, and falls off as 
$r^{-4}$ in the outer envelope ($r > r_{_0}$). We set $\rho_{_0} = 3 
\times 10^{-18}\,{\rm g\,cm}^{-3}$, $r_{_0}=5,000\,{\rm AU}$, and 
$r_{_{\rm B}} = 50,000\,{\rm AU}$. The core mass is then 5.4 M$_{\sun}$. 
We impose a low level of turbulence ($\alpha_{_{\rm TURB}} 
\equiv E_{_{\rm TURB}} / |E_{_{\rm GRAV}}| = 0.05$).

The collapse of the turbulent core leads to the formation of a single star 
surrounded by a disc, and the material in the disc then slowly accretes 
onto the star. A sink particle, representing the star and the inner
part of the disc, is created where the density first exceeds 
$\rho_{_{\rm CRIT}} = 10^{-11}\,{\rm g\,cm}^{-3}$.
We use a newly developed type of sink called a {\it smartie}, which is a 
rotating oblate spheroid. The star is a point mass at the centre of 
the smartie.The smartie gains mass from the accreting material and can lose mass
by ejecting material in the form of protostellar jets. The jets are simulated by
SPH particles which are created at rate $0.1 TM_{_{\rm S}} / t_{_{\rm VISC}}$ 
and launched along the rotation axis with speed $100\,{\rm km\,s}^{-1}$ to form a 
bipolar outflow. These jets push the surrounding material aside and create 
hourglass cavities about the smartie rotation axis (see Stamatellos et al. 2005 for
details).

\subsection{Radiative transfer}

We perform radiative transfer simulations on 3 time-frames during the collapse 
of a star-forming core (see Table~\ref{tab:model.params}). 
We focus our attention just before and just after the 
formation of the first protostar in the core.

For the radiative transfer simulations, we use a version of {\sc Phaethon}, 
a Monte Carlo radiative transfer code which we have developed 
and optimised for radiative transfer simulations on SPH 
density fields. The main new features of the optimised version 
are (i) that it uses the tree structure inherent within the SPH code to 
construct a grid of cubic radiative transfer (RT) cells, spanning the 
entire computational domain, the {\it global grid}; and (ii) that in 
addition it constructs a local grid of RT cells 
around each star, a {\it star grid}, representing a flared disc,
to capture the steep temperature gradients that are expected in the vicinity of a star
(Stamatellos \& Whitworth 2005).

The core is illuminated externally by the interstellar radiation field and 
internally by the newly formed protostar (once it has formed).
For the external radiation field  we adopt a revised version of the Black 
(1994) interstellar radiation field, which consists 
of radiation from giant stars and dwarfs, thermal emission from dust grains, 
cosmic background radiation, and mid-infrared emission from transiently heated 
small PAH grains (Andr\'e et al. 2003).
The protostar luminosity is dominated by the luminosity  due to accretion since
the mass of the young protostar is small ($<0.6\,{\rm M}_{\sun}$) 
and the accretion rate is high ($>10^{-7}\,{\rm M}_{\sun}$/yr).

\begin{figure}
\centerline{
\includegraphics[width=9.1cm]{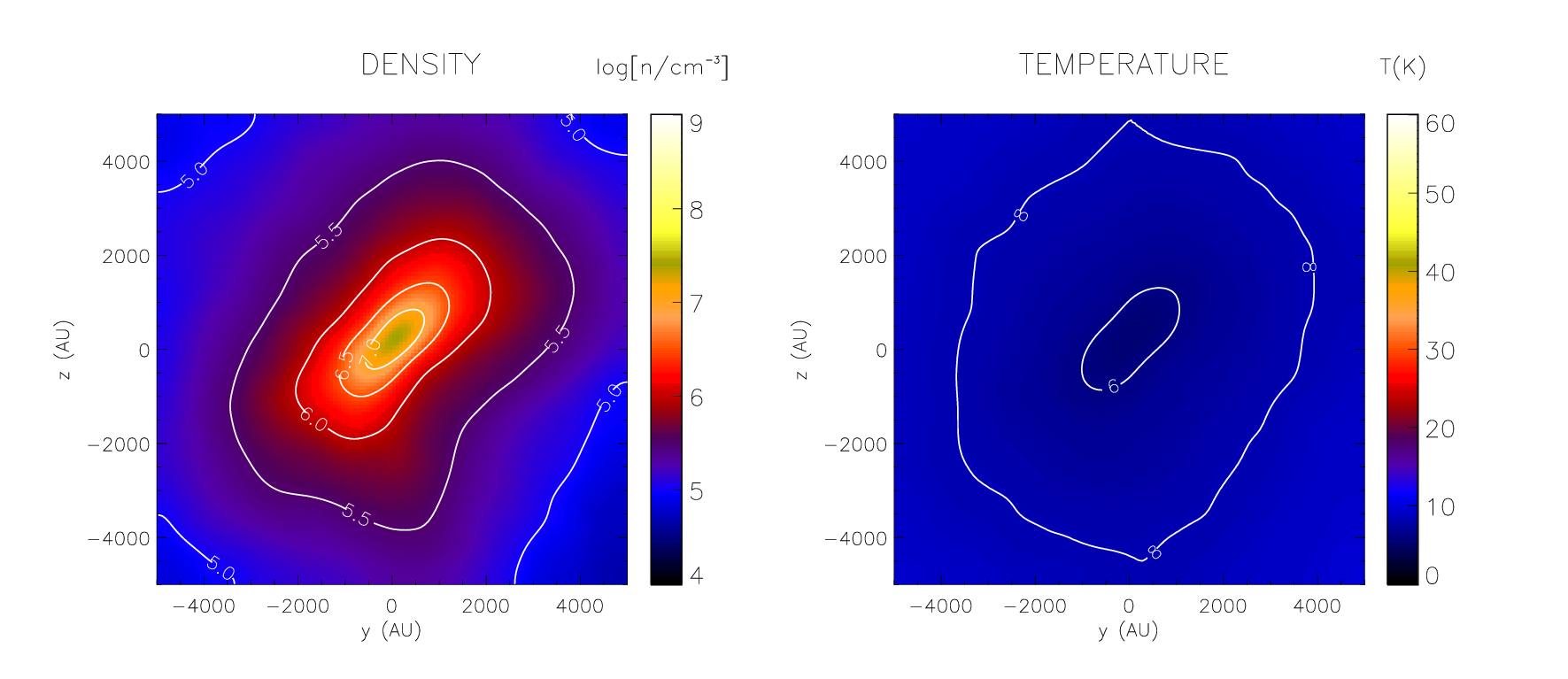}}
\centerline{
\includegraphics[width=9.1cm]{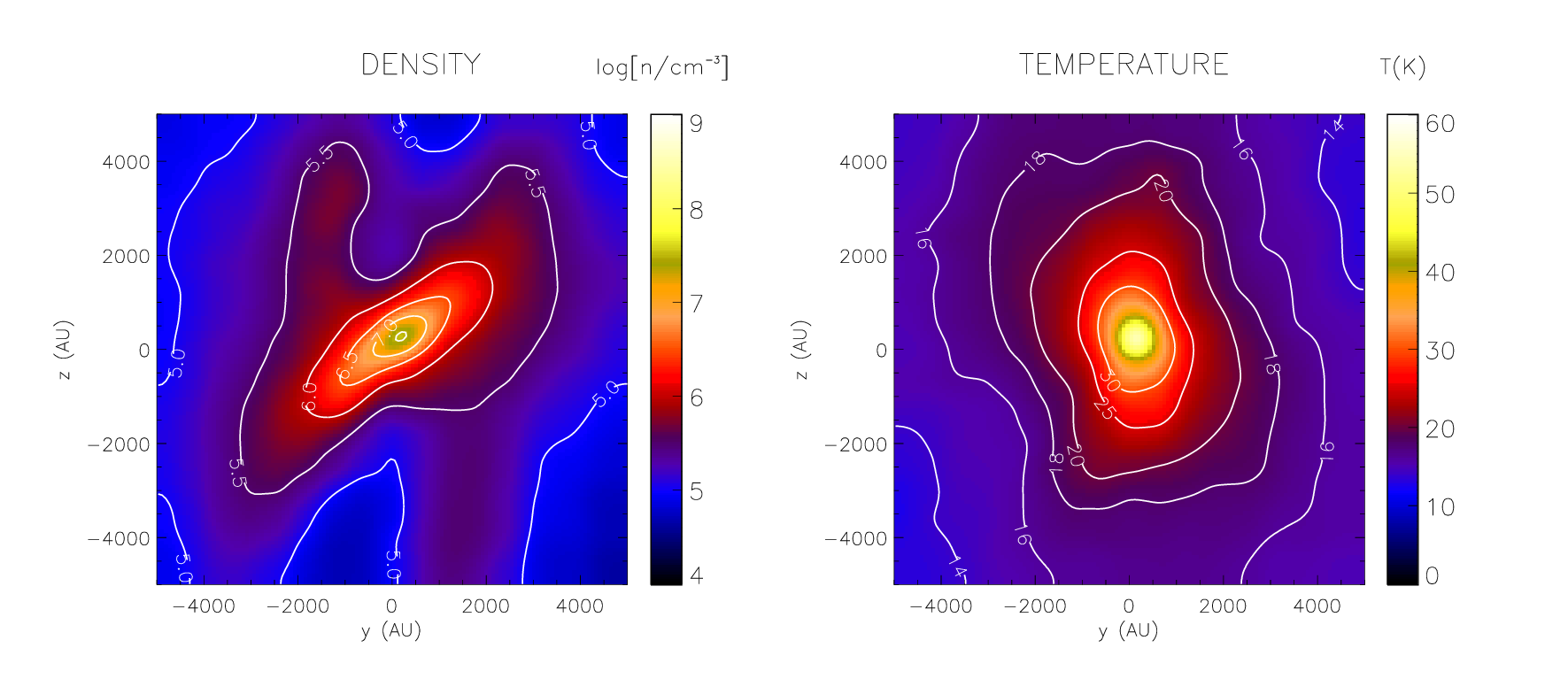}}
\centerline{
\includegraphics[width=9.1cm]{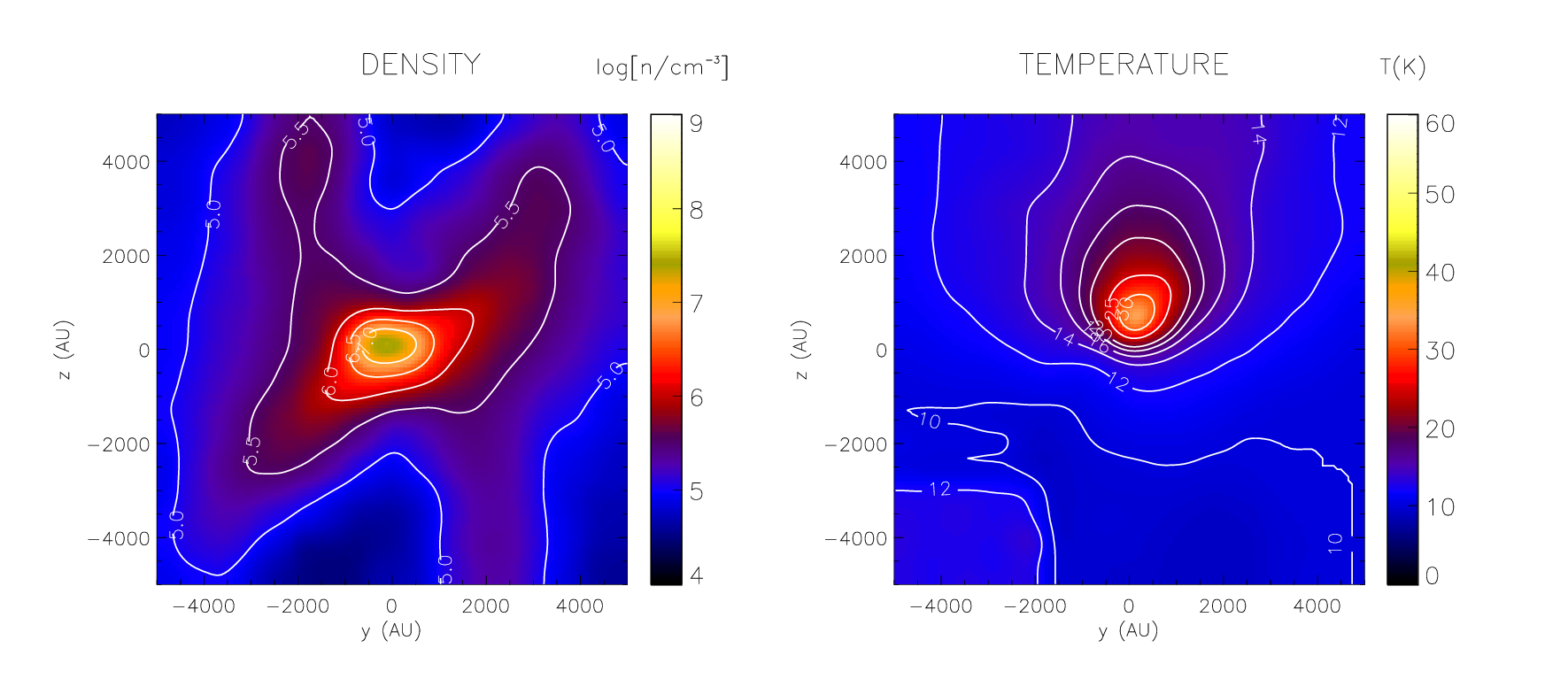}}
\caption{Cross sections of density and dust temperature on a plane 
parallel to the $x=0$ plane of 3 different time-frames (first row: 
{\it collapsing prestellar core};$\;$ second row: 
{\it Class 0 object};$\;$ third row: {\it Class 0 object}).} 
\label{fig.dens.temp}
\end{figure}

\begin{figure*}
\centerline{
\includegraphics[width=4.5cm]{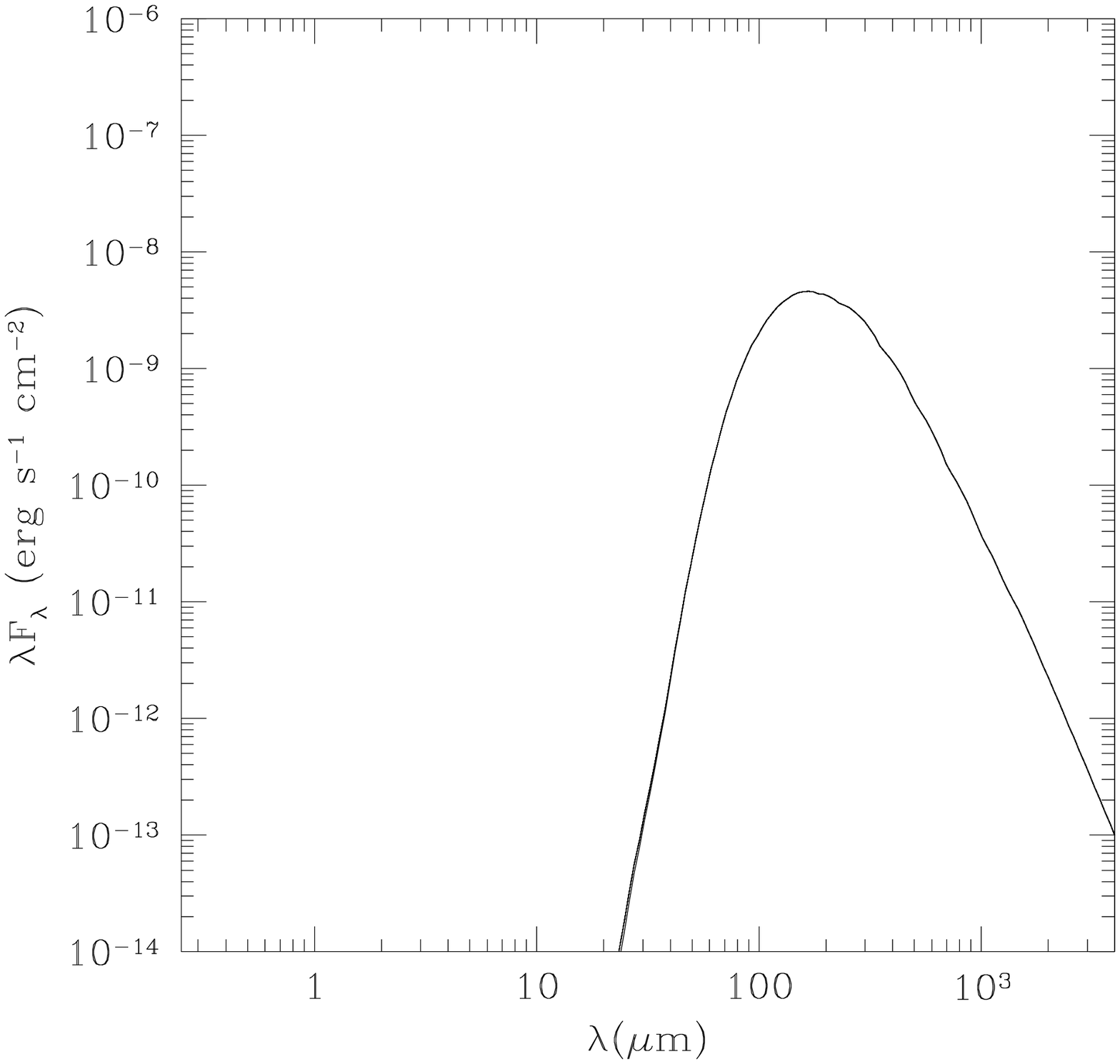}\hspace{1.5cm}
\includegraphics[width=4.5cm]{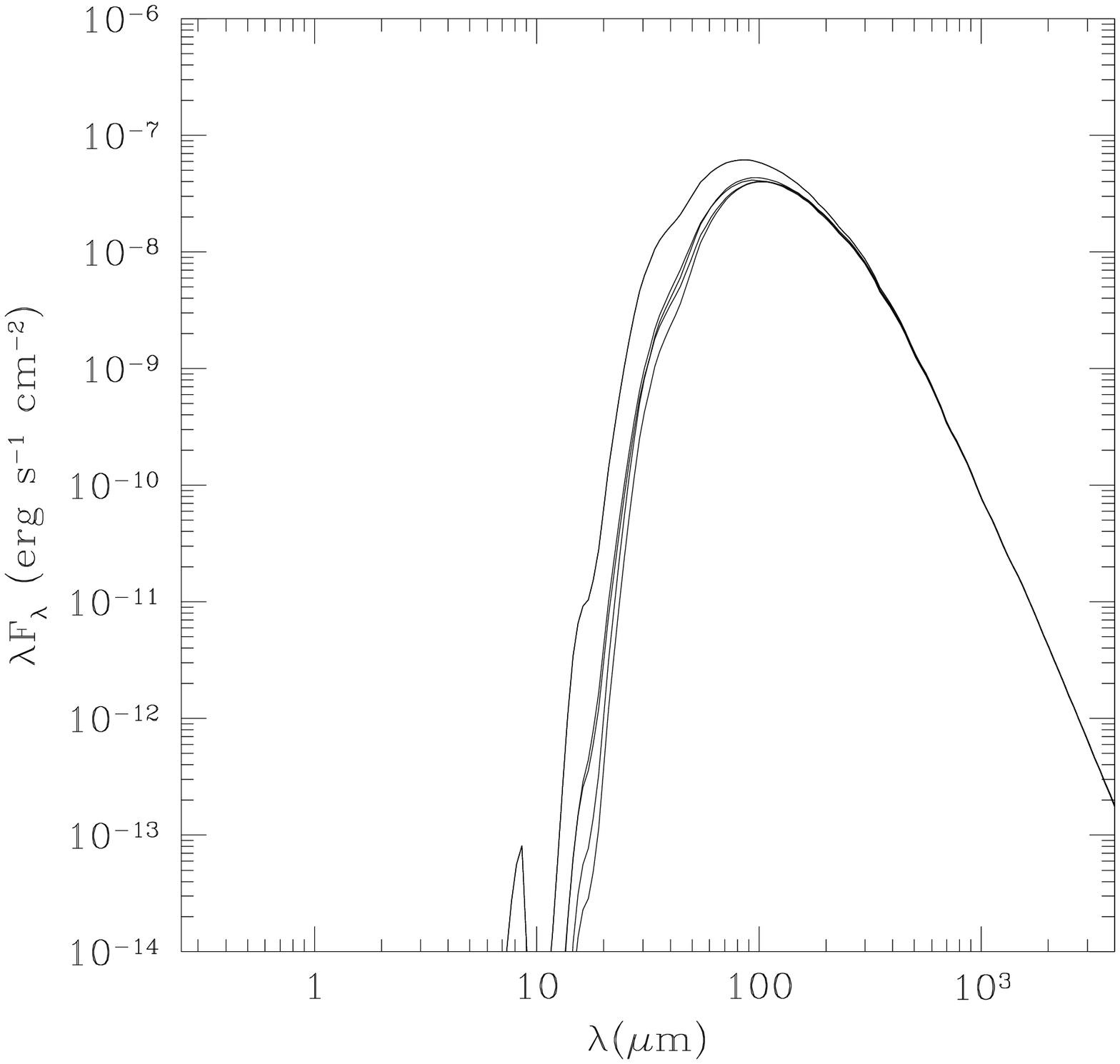}\hspace{1.5cm}
\includegraphics[width=4.5cm]{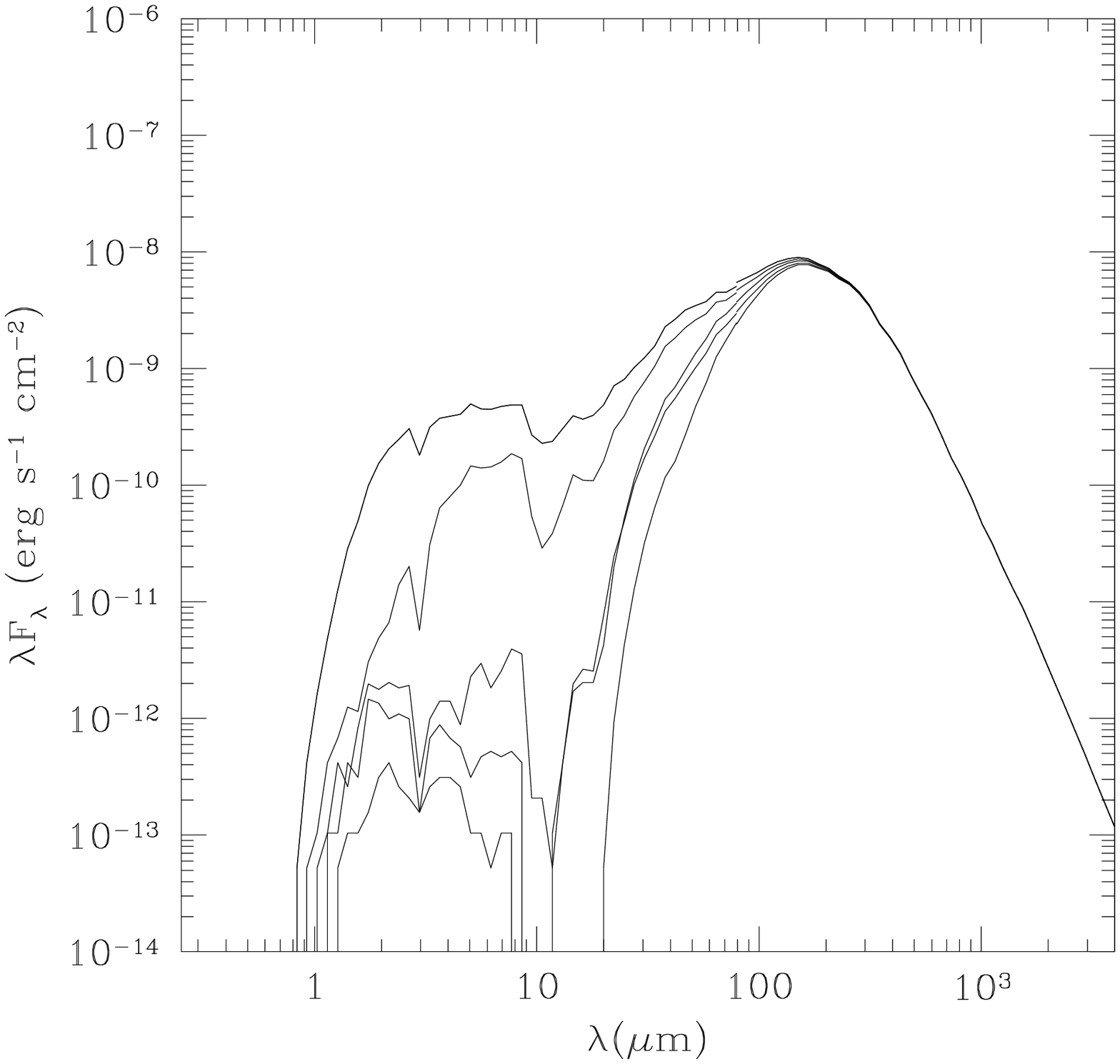}
}\vspace{-0.2cm}
\caption{SEDs of 3 different time-frames  from 6 different angles
(the SED is weakly dependent on viewing angle so that some of the curves
overlap; the flux is higher when the protostar is viewed pole-on, i.e. through the hourglass cavity that is created by the jets).
{\bf (a)} first column - {\it collapsing prestellar core}. 
The SED peaks at $\sim 190\,\micron$
(temperature $T_{_{\rm EFF}} \sim 13\,{\rm K}$).
{\bf (b)}  second column - {\it Class 0 object}.
The emission of the system peaks at $\lambda \sim 
80\;{\rm to}\,100\,\micron$ (depending on the observer's  viewing angle)
($T_{_{\rm EFF}} \sim 25\,{\rm to}\,31\,{\rm K}$). 
{\bf (c)} third column - {\it Class 0 object.} 
The protostar has moved out of the central region by $\sim 100\,{\rm AU}$, and 
the attenuated stellar emission can be seen at short wavelengths ($1\;{\rm to}\,50\,\micron$).The peak of the cloud emission is at $\lambda \sim 150\,\micron$ ($T_{_{\rm EFF}}
\sim 17\,{\rm K}$).}
\label{fig.seds}
\end{figure*}

\begin{figure*}
\centerline{
\includegraphics[width=6.1cm]{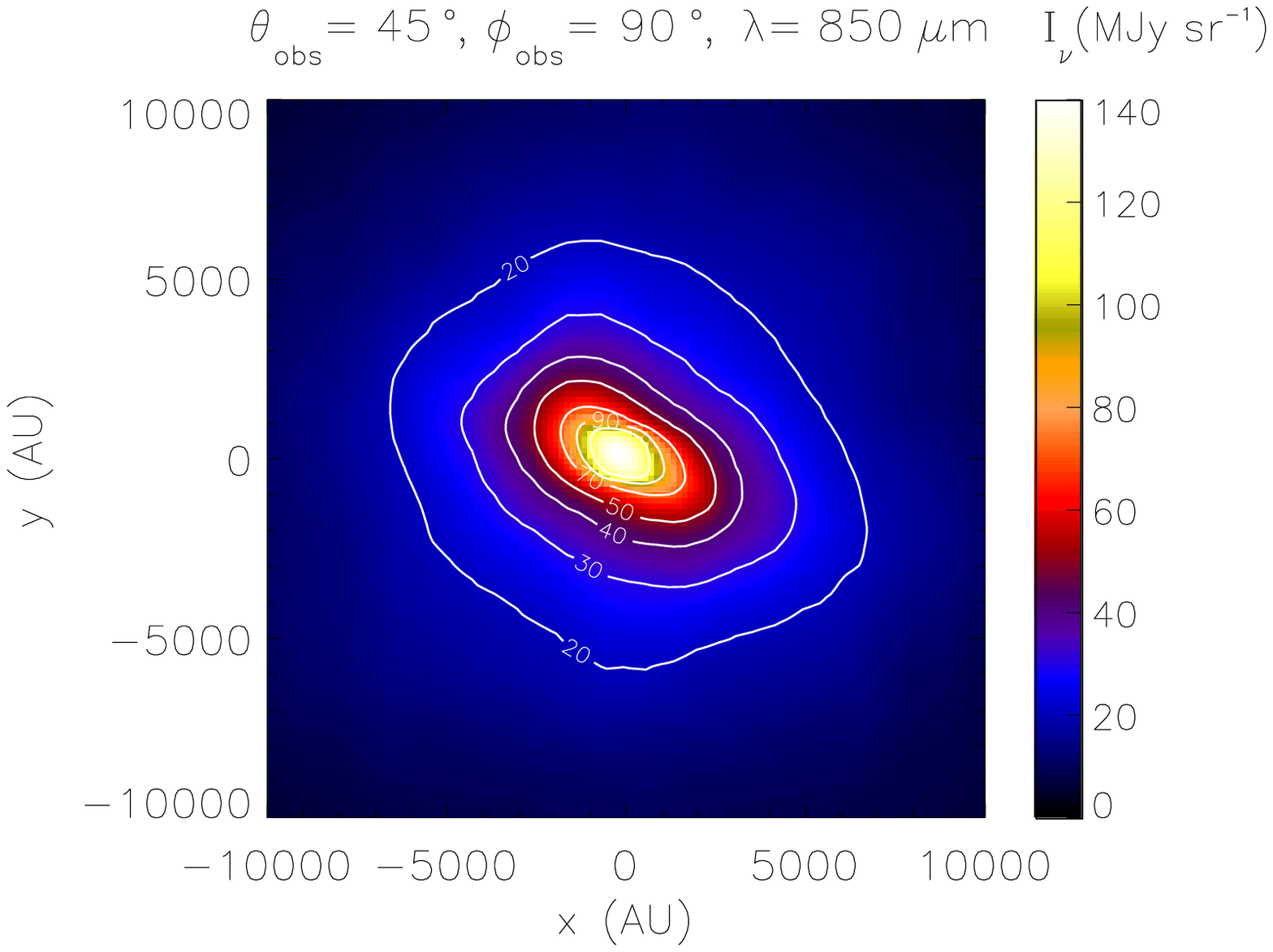}\hspace{-.3cm}
\includegraphics[width=6.1cm]{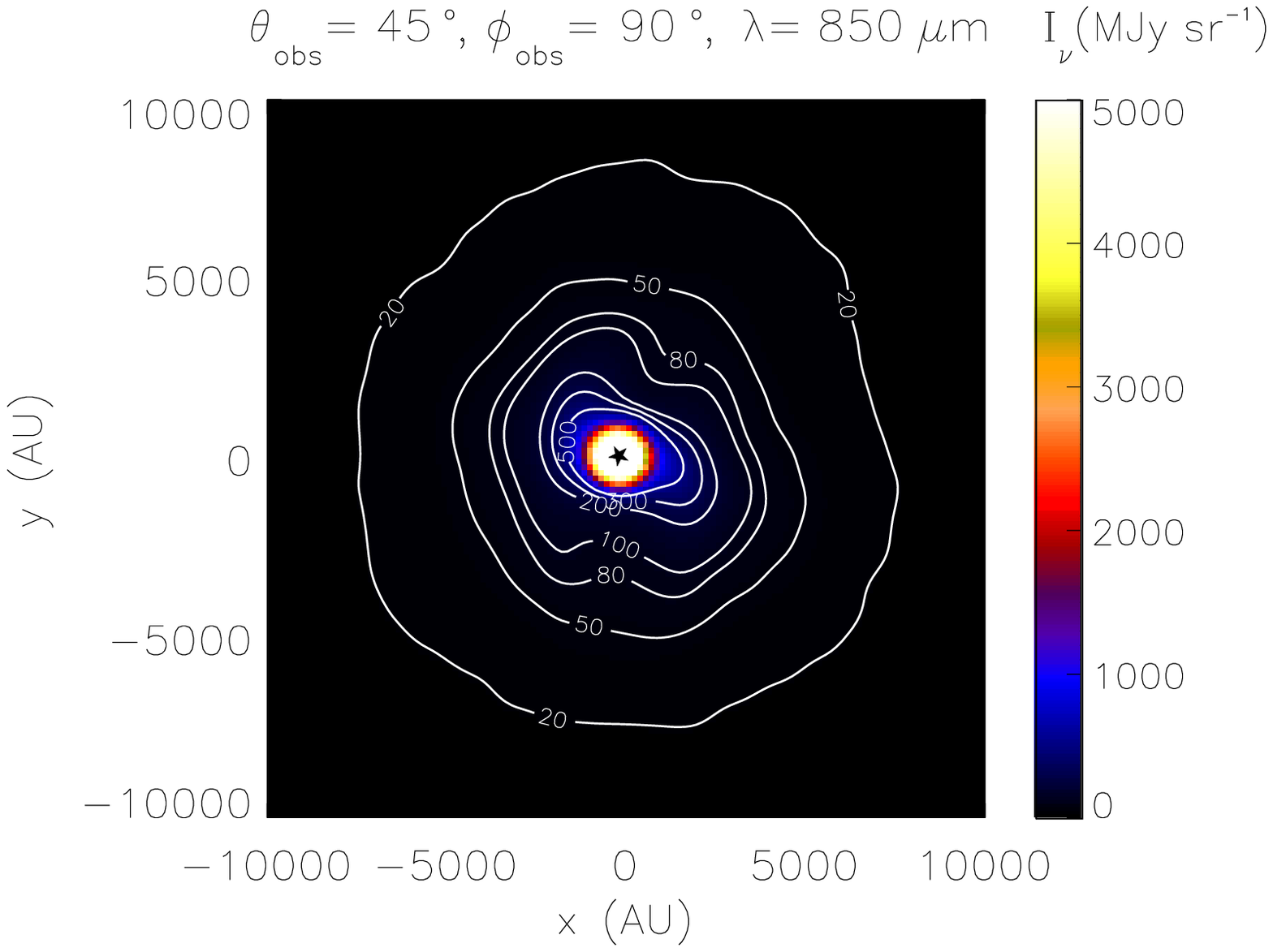}\hspace{-.3cm}
\includegraphics[width=6.1cm]{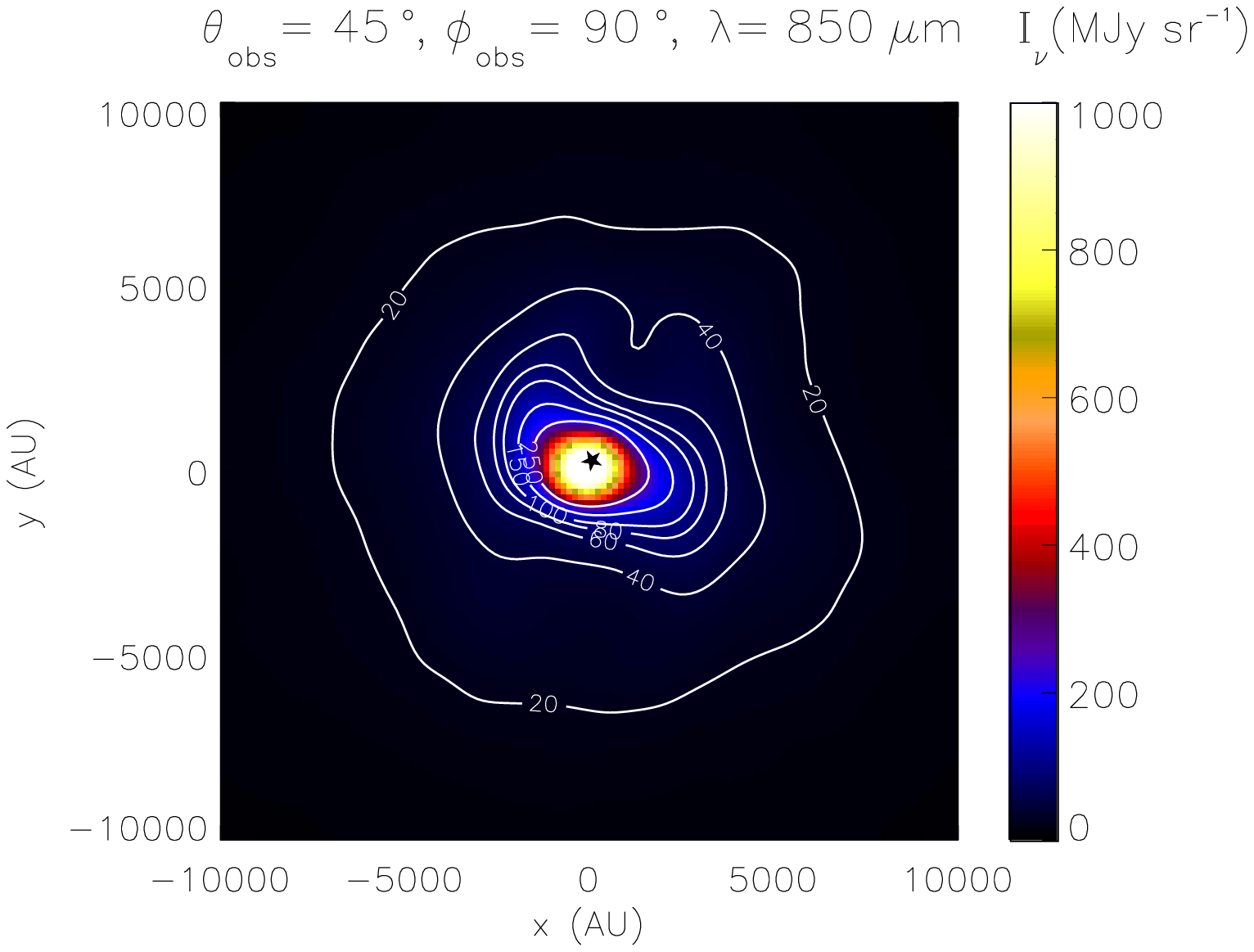}
}
\caption{850 $\micron$ isophotal 
maps of 3 different time-frames (first column: 
{\it collapsing prestellar core};$\;$ second column: 
{\it Class 0 object};$\;$ third column: {\it Class 0 object}).
Cores with protostars (second and third columns) 
are more centrally condensed than prestellar cores (first column).
The axes $(x,y)$ refer to the plane of sky 
as seen by the observer}
\label{fig.isomaps}
\end{figure*}

\begin{figure*}
\centerline{
\includegraphics[width=6.1cm]{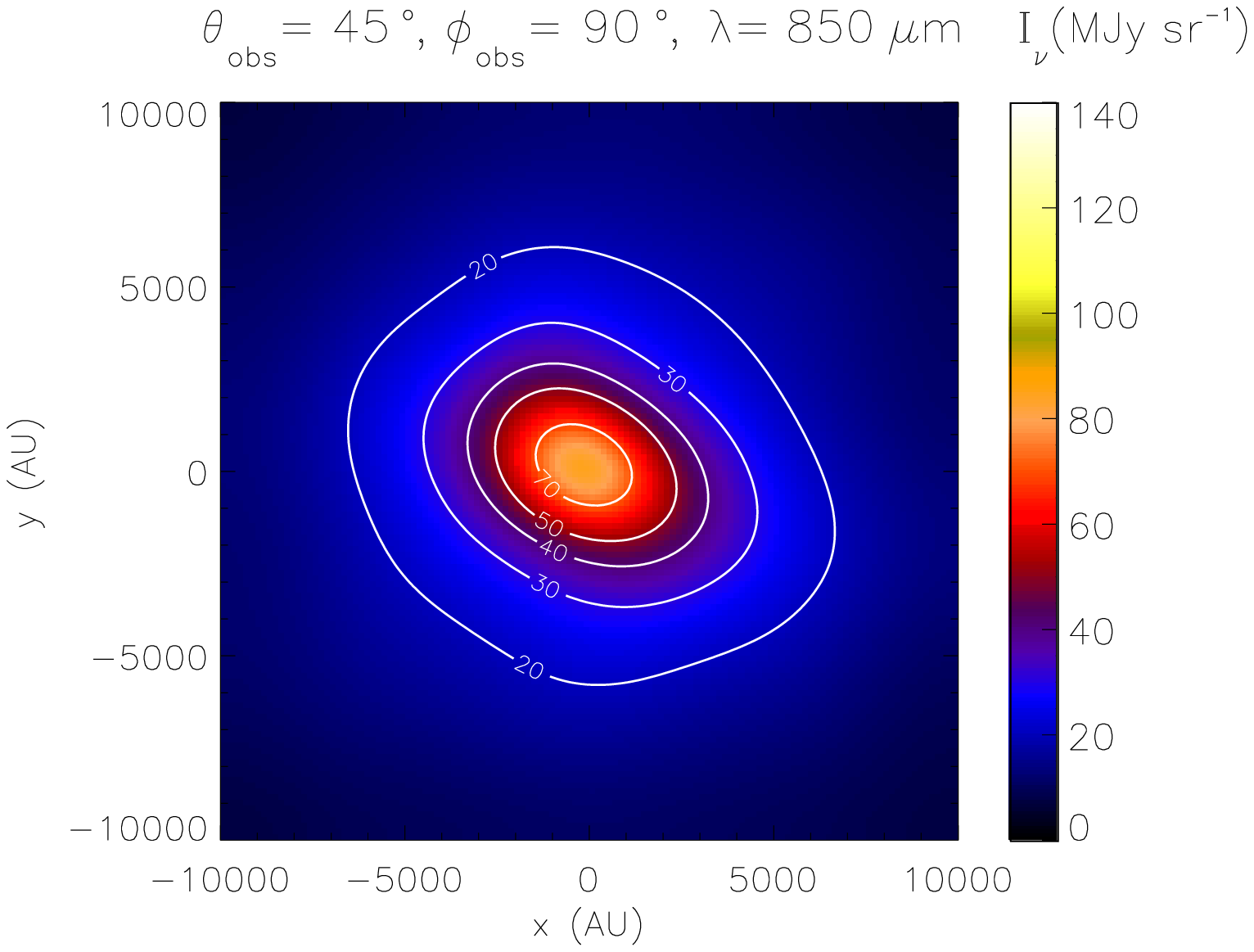}\hspace{-.3cm}
\includegraphics[width=6.1cm]{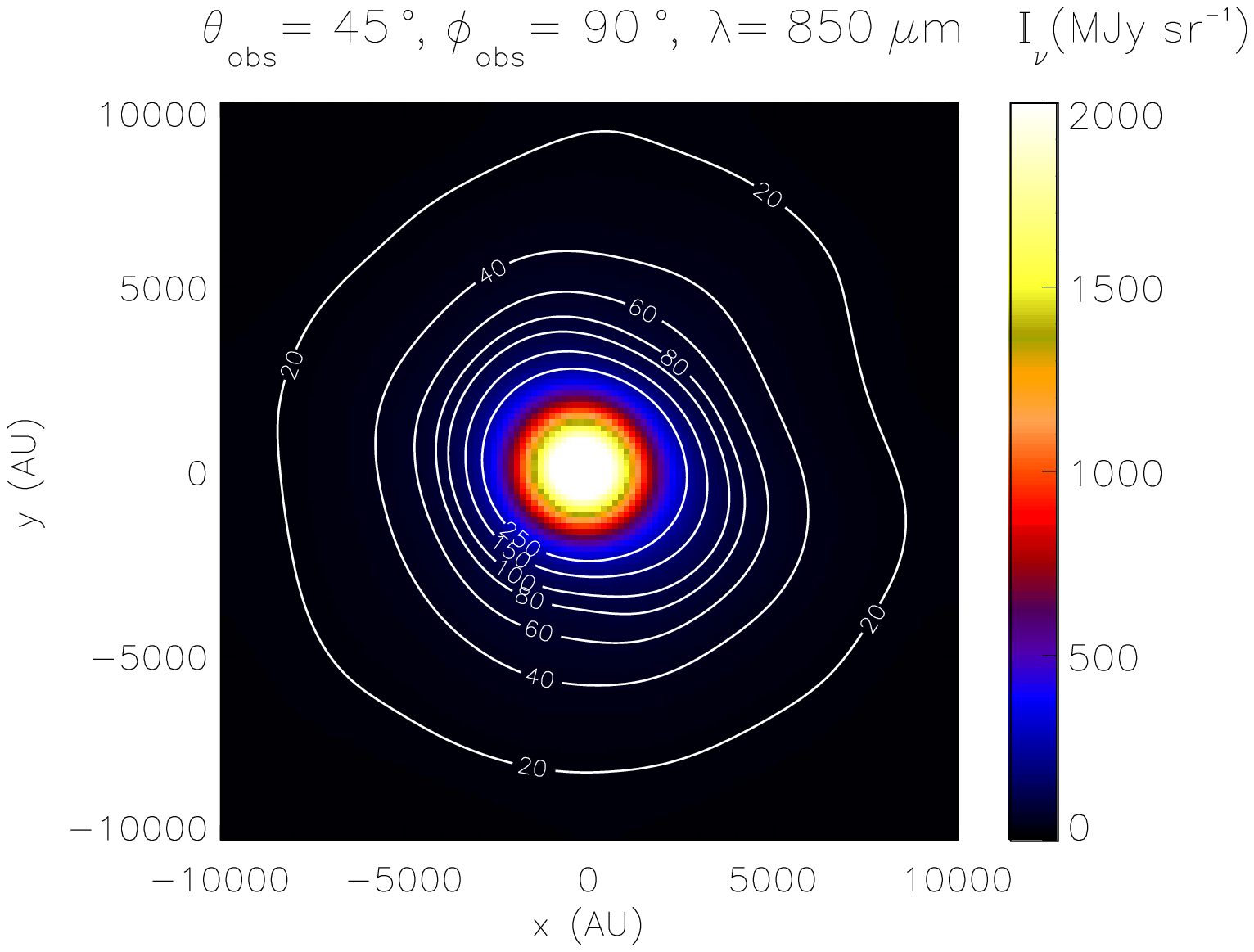}\hspace{-.3cm}
\includegraphics[width=6.1cm]{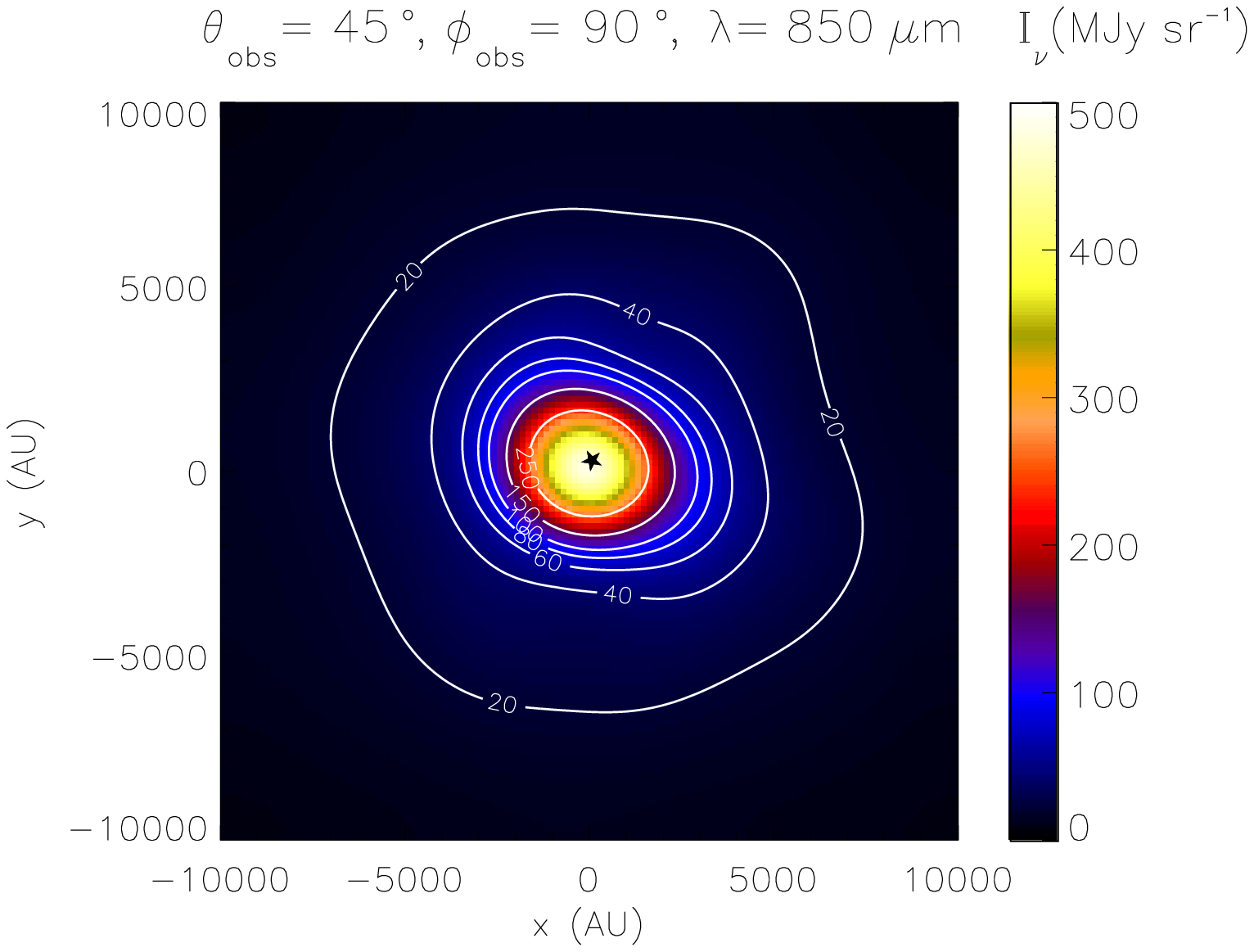}
}
\caption{As in Fig.~\ref{fig.isomaps}, but after convolving with 
a Gaussian beam (${\rm FWHM} =  2300\,{\rm AU}$, e.g.
$15\arcsec$ for a core at $150\,{\rm pc}$). Cores with protostars appear more circular, 
but otherwise similar to the prestellar core.}
\label{fig.isomaps.conv}
\end{figure*}

\section{Characteristics of prestellar cores and young protostars}

\subsection{Dust temperatures}

In Fig.~\ref{fig.dens.temp} we present the density fields and dust-temperature 
fields for the 3 time-frames in Table~\ref{tab:model.params}. We plot the density 
and the temperature on a cross section through the computational domain parallel to 
the $x=0$. This plane passes 
through the protostar, or, if there is no protostar (as in the first time-frame), 
through the densest part of the core. The plots show only the central region of the 
core ($5000\,{\rm AU} \times 5000\,{\rm AU}$). 

Our results for the temperature are broadly similar to those of previous 1D and 
2D studies of prestellar cores and Class 0 objects
(Evans et al. 2001; Zucconi et al. 2001; Shirley et al. 2002; Stamatellos \& Whitworth 2003).
The prestellar core is quite cold ($5\;{\rm to}\,20\,{\rm K}$). 
As soon as a protostar forms, the region around it 
becomes very hot (up to the dust destruction temperature), but the temperature 
drops below $\sim\!20\,{\rm K}$ beyond a few $1000\,{\rm AU}$ from the protostar, 
because of the high optical depth in the dense accretion flow onto the protostar.

\subsection{SEDs}

The SEDs of the 3 time-frames in Table~\ref{tab:model.params} are presented in
Fig.~\ref{fig.seds}. These SEDs have been calculated assuming that the core is 
at $140\,{\rm pc}$. SEDs are plotted for 6 different viewing angles, i.e. 3 polar 
angles ($\theta=0\degr$, $45\degr$, $90\degr$) and 2 azimuthal angles 
($\phi=0\degr$, $90\degr$). 

The effective temperature of the core, as inferred from the peak of the SED, rises 
and falls with the accretion luminosity of the protostar. For a prestellar core, 
the SED peaks at $\sim 190\,\micron$, implying $T_{_{\rm EFF}} \sim 13\,{\rm K}$. 
Once the protostar has formed and inputs energy into the system, the peak moves 
steadily to shorter wavelengths, reaching $\sim 80\,\micron$ ($T_{_{\rm EFF}} \sim 
31\,{\rm K}$) as the accretion luminosity reaches its maximum, and then moving back 
to longer wavelengths again as the accretion luminosity declines. By the final frame 
it has reached $\sim 150\,\micron$ ($T_{_{\rm EFF}} \sim 17\,{\rm K}$).

The peak of the SED of a prestellar core is independent of viewing angle, since 
the core is optically thin to the radiation it emits. In contrast, the peak of the 
SED of a Class 0 object does depend on viewing angle, albeit weakly, because of the 
presence of an optically thick disc around the protostar. However, even allowing for 
variations in the viewing angle, the SED of a Class 0 object does not peak at the 
wavelengths characteristic of prestellar cores ($\sim 190\,\micron$).

Our model predicts that a young protostar embedded in a core is not observable in the NIR,
{\it unless} it is displaced from the central high-density region. This refers to the very
early stages of protostar formation. This result contradicts the recent results of 
Whitney et al. (2003) and Young et al. (2004); they predict that 
NIR radiation from a deeply embedded young protostar is observable.
We attribute the differences to the fact that these models use different densities for
 the region near the protostar and different 
opacities. Ultimately, the density distribution within 500 AU of the the centre of a core or a
protostar is not well constrained, and therefore radiative transfer calculations
have to rely on theoretical models of core collapse. Dust opacities are also poorly 
constrained (e.g. Bianchi et al. 2003).

\section{Isophotal maps}

In Fig.~\ref{fig.isomaps} we present isophotal maps at $850\,\micron$ for the 
3 time-frames in Table 1.
At $850\,\micron$ the core is optically thin, 
and the temperature does not vary much ($\sim 10\;{\rm to}\,20\,{\rm K}$, 
apart from the region very close to the protostar in time-frames {\texttt 
t3} and {\texttt t5}), so the maps are effectively column density maps. 
Class 0 objects are more centrally condensed than prestellar cores. They 
also show more structure, due to bipolar outflows, which clear low-density 
cavities.

In Fig.~\ref{fig.isomaps.conv} we present the isophotal maps after 
convolving them with a Gaussian beam having ${\rm FWHM} = 2,300\,{\rm AU}$. 
Assuming the cloud is 150~pc away, this corresponds to an angular resolution 
of $15\arcsec$, which is similar to the beam size of current submm and mm 
telescopes (e.g. SCUBA, IRAM). On these maps, Class 0 objects look very 
similar to prestellar cores, apart from the fact that they tend to appear 
more circular and featureless.

The reason for this is that the emission from a core that contains a protostar is 
concentrated in the central few hundred AU, and so, when it is convolved with 
a $2,300\,{\rm AU}$ beam, it produces an image rather like the beam, i.e. 
round and smooth. In contrast, the emission from a prestellar cores has structure 
on scales of several thousand AU, much of which survives convolution with a 
$2,300\,{\rm AU}$ beam. Thus we should expect cores that contain protostars to 
appear rounder than prestellar cores, and indeed this is what is observed 
(Goodwin et al. 2002).

\section{Conclusions}

Our simulations indicate that a newly-formed protostar may not be observable in the NIR 
with current telescopes, because it is too deeply embedded. 
This result reflects the high densities which our model predicts in the 
immediate surroundings of a newly-formed protostar. Thus, it is suggested that
cores with {\it no observed} NIR radiation may in fact contain the {\it youngest} protostars.

Based on the results of the combined hydrodynamic and radiative transfer simulations, we propose
two criteria for identifying cores which
 -- despite appearing to be prestellar -- may harbour very young protostars:

(a) They are warm ($T>15\,{\rm K}$) as indicated by the peak of the SED of the 
core ($\lambda_{\rm peak} < 170\,\micron$). This criterion requires that the peak 
of the SED can be measured to an accuracy of $\sim 30\,\micron$, which should be 
feasible with observations in the far-IR from ISO 
and the upcoming Herschel mission, and in the mm/submm region from 
SCUBA and IRAM. We have presumed that the 
core is heated by the average interstellar radiation field, and the criterion 
would have to be modified for a core irradiated by a stronger or weaker radiation 
field.

(b) Their submm/mm isophotal maps are circular, at least in the central 
$2000\,{\rm to}\,4000\,{\rm AU}$. 

These criteria can be used to identify ``warm'', circular cores that may potentially
contain the youngest protostars. These cores can then be probed by deep radio observations 
with the VLA. Recently, we carried out 3.6 cm VLA observations for 4 of these cores 
and results will be presented in a future publication.

\begin{acknowledgements}
We thank P. Andr\'e for providing the revised version of the Black (1994) ISRF, that accounts
for the PAH emission. 
We also thank D. Ward-Thompson for useful discussions and suggestions. 
This work was partly supported
from the EC Research Training Network ``The Formation and Evolution of
Young Stellar Clusters'' (HPRN-CT-2000-00155), and partly by PPARC grant 
PPA/G/O/2002/00497.
\end{acknowledgements}

\end{document}